\documentclass[prl,twocolumn,reprint]{revtex4-1}
\usepackage{graphicx}
\usepackage{dcolumn}
\usepackage{bm}
\usepackage{color}
\usepackage{tabularx}
\usepackage{array}
\usepackage{amsmath}
\usepackage{lipsum}

\begin{document}

\title{Characterization of spin relaxation anisotropy in Co using spin pumping}

\author{Y. Li, W. Cao and W. E. Bailey}

\affiliation{Materials Science \& Engineering, Department of Applied Physics and Applied Mathematics, Columbia University, New York NY 10027, USA}

\date{\today}

\begin{abstract}

Ferromagnets are believed to exhibit strongly anisotropic spin relaxation, with relaxation lengths for spin longitudinal to magnetization significantly longer than those for spin transverse to magnetization. Here we characterize the anisotropy of spin relaxation in Co using the spin pumping contribution to Gilbert damping in noncollinearly magnetized Py$_{1-x}$Cu$_{x}$/Cu/Co trilayer structures. The static magnetization angle between Py$_{1-x}$Cu$_{x}$ and Co, adjusted under field bias perpendicular to film planes, controls the projections of longitudinal and transverse spin current pumped from Py$_{1-x}$Cu$_{x}$ into Co. We find nearly isotropic absorption of pure spin current in Co using this technique; fits to a diffusive transport model yield the longitudinal spin relaxation length $< 2$ nm in Co. The longitudinal spin relaxation lengths found are an order of magnitude smaller than those determined by current-perpendicular-to-planes giant magnetoresistance measurements, but comparable with transverse spin relaxation lengths in Co determined by spin pumping.

\end{abstract}

\maketitle

\indent A key question for spin electronics concerns the relaxation mechanisms for spin current injected into a variety of materials. Spin relaxation in ferromagnets (Fs), central for spin momentum transfer, is special because of the anisotropy axis presented by the spontaneous magnetization $\mathbf{M}$ \cite{valetPRB1993,zhangPRL2002,stilesPRB2002,tserkovnyakPRL2002,heinrichPRL2003,ShpiroPRB2003,ZhangjwPRB2004,ZhangjwPRL2004,ZhangjwPRB2005,petitjeanPRL2012}. Longitudinal spin relaxation\cite{valetPRB1993}, with spin polarization $\sigma$ parallel (antiparallel) to $\mathbf{M}$, causes spin accumulation to decrease exponentially with distance over a scale greater than the electronic mean free path\cite{zhangPRL2002}. Transverse spin relaxation, with $\sigma$ orthogonal to $\mathbf{M}$, is governed by the dephasing process of spin-up and spin-down eigenmodes due to their different Fermi wavevectors, leading to oscillation and decay of spin accumulation on a scale shorter than the electronic mean free path\cite{stilesPRB2002,petitjeanPRL2012}. \\
\indent The characteristic length scales for the two different spin relaxation processes in ferromagnets, $\lambda_\text{sr}^L$ for longitudinal and $\lambda_\text{sr}^T$ for transverse spin relaxation, have been evaluated largely using two separate experimental techniques: magnetotransport\cite{pirauxEPJB1998} for $\lambda_\text{sr}^L$ and ferromagnetic resonance (FMR)\cite{tserkovnyakPRL2002,heinrichPRL2003} for $\lambda_\text{sr}^T$. These two measurements characterize charge-accompanied and chargeless spin current, respectively\cite{valetPRB1993,tserkovnyakPRL2002}. Estimates of $\lambda_\text{sr}^L$ come from the F layer thickness dependence of current-perpendicular-to-planes giant magnetoresistance (CPP-GMR)\cite{pirauxEPJB1998,steenwykJMMM1997,duboisPRB1999, jbassJPCM2007}; extracted values of $\lambda_\text{sr}^L$ range from 5 nm for Ni$_{79}$Fe$_{21}$ up to 40 nm for Co at room temperature. FMR measurements of spin pumping, for collinearly magnetized F$_1$/Cu/F$_2$ structures, show much shorter penetration depths ($\lambda_\text{C}$) to fully absorb transverse spin current\cite{ghoshPRL2012,taniguchiAPEX2008}. Co has the most anisotropic spin relaxation according to these separate measurements, with $\lambda_\text{sr}^L/\lambda_\text{sr}^T\sim$16 taking $\lambda_\text{sr}^T\sim 2\lambda_\text{C}=2.4$ nm\cite{ghoshPRL2012, forosJAP2005}. \\
\indent In this manuscript, we demonstrate that the longitudinal spin relaxation length, in addition to the transverse spin relaxation length\cite{ghoshPRL2012}, can also be characterized using a spin pumping measurement, enabling a measurement of the anisotropy of spin relaxation in a given ferromagnetic layer. We present FMR measurements of the spin pumping contribution to Gilbert damping in noncollinearly magnetized Py$_{1-x}$Cu$_{x}$/Cu/Co multilayers (Py=Ni$_{79}$Fe$_{21}$). Using Py$_{1-x}$Cu$_{x}$ alloys, which have adjustably smaller saturation magnetization $M_\text{s}$ than Co, we can change the magnetization alignment of Py$_{1-x}$Cu$_{x}$ and Co from collinear for in-plane FMR to near-orthogonal for perpendicular FMR. As the angle $\theta_\text{M}$ between Py$_{1-x}$Cu$_{x}$ and Co magnetization tends towards $\pi/2$, one component of injected spin from Py$_{1-x}$Cu$_{x}$ tends towards the longitudinal direction (Fig. 1), allowing us to probe anisotropy in spin relaxation through the linewidth of the Py$_{1-x}$Cu$_x$ layer\cite{tserkovnyakPRB2003,taniguchiPRB2007}. We find, surprisingly, that spin relaxation, as measured through the spin pumping contribution to Gilbert damping, is mostly isotropic. In our Co films we estimate $\lambda_\text{sr}^L < 2$ nm for all different Py$_{1-x}$Cu$_{x}$/Cu/Co samples, which is comparable to its transverse counterpart($\sim$2.4 nm) but inconsistent with the much longer ($\sim$40 nm) lengths reported from room-temperature CPP-GMR\cite{pirauxEPJB1998,jbassJPCM2007}.\\
\begin{figure}[htb]
\centering
\includegraphics[width = 2.5 in] {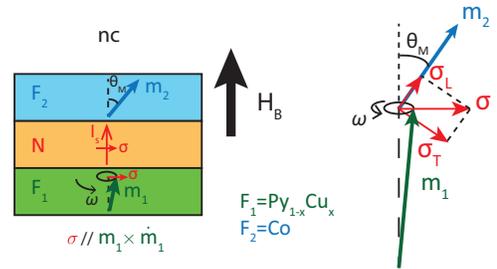}
\caption{\textit{Left:} Noncollinear magnetization alignment of the F$_1$/N/F$_2$ trilayer at the FMR condition for F$_1$. \textit{Right:} $\mathbf{m}_1$ is driven into precession, pumping spin current into $\mathbf{m}_2$, with spin components both longitudinal ($\sigma_\text{L}$) and transverse ($\sigma_\text{T}$) to the $\mathbf{m}_2$ magnetization.}
\label{fig1}
\end{figure}
\indent Three types of thin-film heterostructures were prepared by UHV sputtering and characterized by FMR. Pseudo-spin-valve-type Py$_{1-x}$Cu$_{x}$(t)/Cu(5 nm)/Co(5 nm) trilayers were used to characterize the anisotropy of spin-current absorption in Co. Their response was compared with two types of Py$_{1-x}$Cu$_{x}$(t) alloy control samples. Bilayers of Py$_{1-x}$Cu$_{x}$(5 nm*)/Cu(5 nm) and trilayers of Py$_{1-x}$Cu$_{x}$(5 nm*)/Cu(5 nm)/Pt(3 nm) were used to characterize the background damping of the alloy and the spin mixing conductance of the alloy/Cu interface, respectively.  Co(5 nm)/Cu(5 nm)/Pt(3 nm) is also deposited. For the alloy Cu contents $x=0$ to 0.4 were prepared in each case, using confocal sputtering from Py and Cu targets\cite{guanJAP2007}; thicker (10 nm*) alloy layers were used for $x=0.4$. All layers were deposited on Si/SiO$_2$ substrates, seeded by Ta(5 nm)/Cu(5 nm) and capped by Ta(2 nm). See Ref. \cite{chengRSI2012} for details on preparation.\\
\indent Room temperature, variable frequency (3-26 GHz), swept-field FMR measurements were used to characterize the samples, with instrumentation as described in \cite{LiPRL2016}. In order to characterize FMR relaxation of the Py$_{1-x}$Cu$_{x}$ layer under noncollinear magnetization alignment with Co, two types of measurements were carried out. First, we compare the frequency-dependent linewidths of Py$_{1-x}$Cu$_{x}$ and Py$_{1-x}$Cu$_{x}$/Cu/Co samples in both in-plane (parallel-condition, pc) and perpendicular (normal-condition, nc) FMR\cite{anglealignment}, for a series of four measurements at a given alloy content $x$; see Figs. \ref{fig2}c, \ref{fig3}, and \ref{fig4}. Here we expect the Co magnetization of trilayer samples to vary from fully perpendicular to the film plane at high biasing field $H_\text{B}$ (high $\omega/2\pi$) to nearly parallel to the film plane at low $H_\text{B}$ (low $\omega/2\pi$), while the Py$_{1-x}$Cu$_{x}$ magnetization is always perpendicular to the film plane. Second, we compare the polar angle-dependent linewidths of Py$_{0.8}$Cu$_{0.2}$ and Py$_{0.8}$Cu$_{0.2}$/Cu/Co samples at a fixed frequency of $\omega/2\pi = 10$ GHz; see Fig. \ref{fig5}. Here we expect the misalignment angle $\theta_\text{M}$ to change from zero to maximum as we rotate the biasing field from in-plane (pc) to out-of-plane (nc). \\
\indent Theoretical models for the spin pumping contribution to damping under noncollinear magnetization alignment of symmetric F$_1$/N/F$_1$ structures were developed in Refs. \cite{tserkovnyakPRB2003,taniguchiPRB2007}. We have extended these models to consider asymmetric F$_1$/N/F$_2$ structures where F$_1$=Py$_{1-x}$Cu$_{x}$ and F$_2$=Co in our samples. In the spin valve structure the spin-pumping damping enhancement $\Delta\alpha_\text{sp}$ of F$_1$ is caused by the dissipation of spin current in F$_2$. If F$_1$ and F$_2$ are misaligned by an angle $\theta_\text{M}$, where  $\theta_\text{M}=\mathbf{m}_1\cdot\mathbf{m}_2$ (Fig. \ref{fig1}), during small-angle precession of F$_1$, the polarization of spin current pumped into F$_2$ will oscillate from fully transverse to maximally longitudinal. The instantaneous spin-pumping damping will then oscillate from $\alpha_\text{sp}(0^\circ)=\Delta\alpha_0 \times \tilde{g}_2^{\uparrow\downarrow} / (\tilde{g}_1^{\uparrow\downarrow} + \tilde{g}_2^{\uparrow\downarrow})$, as given in the standard collinear case\cite{tserkovnyakRMP2005}, to a minimum value given by\cite{supplemental}:
\begin{widetext}
\begin{equation}
\Delta \alpha_\text{sp}(\theta_\text{M}) = \Delta\alpha_0 {g_2^*(A\sin^2\theta_\text{M}-B\sin\theta_\text{M}\cos\theta_\text{M})+\tilde{g}_2^{\uparrow\downarrow}(C\cos^2\theta_\text{M}-B\sin\theta_\text{M}\cos\theta_\text{M}) \over AC-B^2}
\label{eq01}
\end{equation}
\end{widetext}
Here $\tilde{g}^{\uparrow\downarrow}_i$ and $g^*_i$ ($i=1$, 2) are the effective transverse and longitudinal spin conductances, respectively; $\Delta\alpha_0=\gamma\hbar \tilde{g}_1^{\uparrow\downarrow}/(4\pi M_st_F)$ is the damping enhancement with effective spin mixing conductance of $\tilde{g}_1^{\uparrow\downarrow}$\cite{LiPRL2016}; in Eq. \ref{eq01} $A(\theta_\text{M}) = g_1^*\sin^2\theta_\text{M}+\tilde{g}_1^{\uparrow\downarrow}\cos^2\theta_\text{M} + \tilde{g}_2^{\uparrow\downarrow}$, $B(\theta_\text{M}) = (\tilde{g}_1^{\uparrow\downarrow} - g_1^*)\sin\theta_\text{M}\cos\theta_\text{M}$ and $C(\theta_\text{M}) = g_1^*\cos^2\theta_\text{M}+\tilde{g}_1^{\uparrow\downarrow}\sin^2\theta_\text{M} + g_2^*$. We take the arithmetic mean of the two extreme cases as the effective damping enhancement, as found to be valid in Ref. \cite{taniguchiPRB2007}. See the Supplemental Materials for details.\\
\indent To maximize the spin pumping anisotropy at finite $\theta_\text{M}$, we use Co (5 nm) for F$_2$, where the dimension is chosen to be significantly thicker than the transverse spin penetration depth, $\lambda_\text{C}=1.2$ nm\cite{ghoshPRL2012}, and thinner than the reported longitudinal relaxation length $\lambda_\text{sr}^L$, $\sim 38$ nm\cite{pirauxEPJB1998}, resulting in a large expected asymmetry in spin relaxation. In the analysis of relaxation in noncollinearly magnetized structures, we take spin mixing conductances $\tilde{g_i}^{\uparrow\downarrow}$ as parameter inputs, determined from the measurements on the Pt control structures, and take the longitudinal spin relaxation length $\lambda^L_\text{sr}$ as a fit parameter. \\
\begin{figure}[htb]
\centering
\includegraphics[width = 3.0 in] {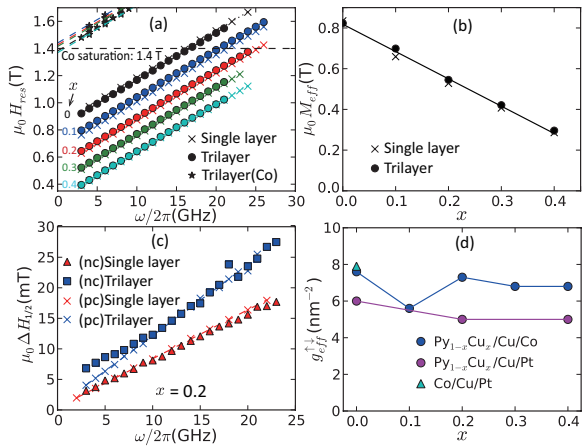}
\caption{(a) Perpendicular (nc-FMR) resonance field $\mu_0H_\text{res}$ for Py$_{1-x}$Cu$_x$ single layers and Py$_{1-x}$Cu$_x$/Cu/Co trilayers, $x=0$ - 0.4, as a function of frequency $\omega/2\pi$. (b) Effective magnetization $\mu_0M_\text{eff}$ extracted from (a) as a function of $x$. (c) Resonance linewidths $\mu_0\Delta H_{1/2}$ of the Py$_{0.8}$Cu$_{0.2}$ single layer and trilayer as a function of frequency $\omega/2\pi$. The spin pumping enhancement is clearly visible in the increased slope ($\alpha$) of the trilayer data; the low-frequency deviation is discussed in Fig. 3. (d) Effective spin mixing conductances $g^{\uparrow\downarrow}_\text{eff}$ of Py$_{1-x}$Cu$_x$/Cu/Co, Py$_{1-x}$Cu$_x$/Cu/Pt and Co/Cu/Pt.}
\label{fig2}
\end{figure}
\indent Fig. \ref{fig2} summarizes the results of fixed-angle nc-and pc-FMR measurements for the three sample series. In Fig. 2(a) we plot resonance fields $\mu_0H_\text{res}$ as a function of frequency for single layers and trilayers in nc-FMR. The good agreement in the $\mu_0 H_\text{res}$ of Py$_{1-x}$Cu$_x$ measured in single layers and trilayers demonstrates that Py$_{1-x}$Cu$_x$ properties are reproducible in deposition. In Fig. 2(b) the effective magnetizations $\mu_0M_\text{eff}$, extracted from fits to the linear Kittel equation $\omega/\gamma = \mu_0(H_\text{res}-M_\text{eff})$, are plotted as a function of $x$. The data show Slater-Pauling dilution of magnetic moment in the Py$_{1-x}$Cu$_x$ layer with increasing Cu content $x$\cite{guanJAP2007}. \\
\indent In Fig. \ref{fig2}(c) we plot full-width half-maximum linewidth $\mu_0\Delta H_{1/2}$ as a function of $\omega/2\pi$ at $x$=0.2. Gilbert-type damping, $\mu_0\Delta H_{1/2} = \mu_0\Delta H_0 + 2\alpha\omega/\gamma$, with negligible inhomogeneous broadening $\mu_0\Delta H_0$, is observed for both pc- and nc-FMR in the single layer and for pc-FMR in the trilayer. The linewidths agree closely for pc- and nc-FMR in the single layer, showing a negligible role for two-magnon scattering in the linewidth\cite{hurbenJAP1998}. In the trilayer, nc and pc linewidths agree well for frequencies above 10 GHz. These observations hold for samples with all Cu content $0\ge x\ge0.4$; the deviations at low frequency are discussed in Fig. \ref{fig3}. The effective spin mixing conductances $g^{\uparrow\downarrow}_\text{eff}$ of trilayer samples are extracted from $\Delta\alpha_\text{sp} = \gamma\hbar g^{\uparrow\downarrow}_\text{eff}/(4\pi M_s t_M)$, shown above where $\Delta\alpha_\text{sp}$ is the difference in $\alpha$ between trilayers and single layers. In Fig. 2(d) we show the extracted $g^{\uparrow\downarrow}_\text{eff}$ for Py$_{1-x}$Cu$_x$/Cu/Co and Py$_{1-x}$Cu$_x$/Cu/Pt structures as a function of $x$. We also plot $g^{\uparrow\downarrow}_\text{eff}$ of Co/Cu/Pt for reference. The lower level of $g^{\uparrow\downarrow}_\text{eff}\sim$7 nm$^{-2}$ for Py$_{1-x}$Cu$_x$/Cu/Co, compared with $\sim$15 nm$^{-2}$ measured in Ref. \cite{ghoshPRL2012}, is likely to be from a more resistive Cu layer, which adds an additional resistance of $(2e^2/h)t_\text{Cu}/\sigma_\text{Cu}$ to the inverse of total spin mixing conductance where $\sigma_\text{Cu}$ is the Cu conductivity. Using these $g^{\uparrow\downarrow}_\text{eff}$ values, we extract the effective spin mixing conductance of Py$_{1-x}$Cu$_x$/Cu and Co/Cu interfaces, shown in the Supplemental Materials\cite{supplemental}. These parameters will be used to determine the longitudinal spin relaxation lengths from the spin pumping data in Figs. \ref{fig4} and \ref{fig5}. \\
\indent In the measurements presented in Fig. \ref{fig2}(c), the nc-FMR linewidths are measured at applied fields below the saturation field for Co, $\mu_0 M_\text{eff}=1.4$ T. The saturation field corresponds to a nc-FMR resonance frequency for Py$_{0.8}$Cu$_{0.2}$ of 25 GHz, as shown in Fig. \ref{fig2}(a). With the resultant noncollinear magnetization alignment in the trilayer, we expect to see spin-pumping damping $\Delta \alpha_\text{sp}$ for Py$_{0.8}$Cu$_{0.2}$ reduced in nc-FMR compared with the values in pc-FMR. Instead, we find that the linewidths of the trilayer measured in pc- and nc-FMR agree closely when $\omega/2\pi>10$ GHz.  Furthermore, in nc-FMR there is an additional broadening from 2-10 GHz in the trilayers which is not predicted by the model. \\
\begin{figure}[htb]
\centering
\includegraphics[width = 3.0 in] {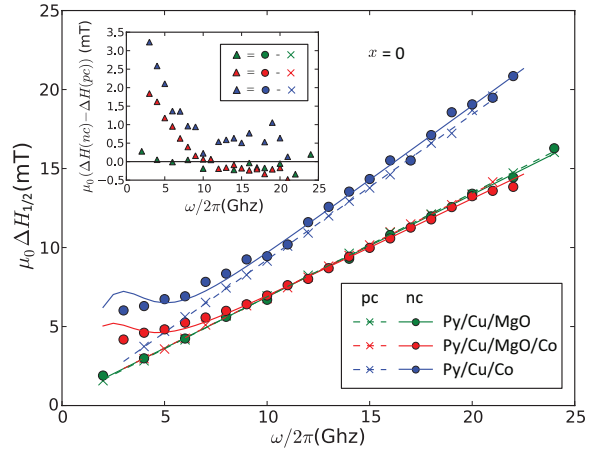}
\caption{
pc- and nc-FMR linewidths for single (Py) and trilayer (Py/Cu/Co) structures, introducing MgO interlayers to suppress spin pumping. Dashed lines are linear fits to pc-FMR linewidths. Solid curves assume (magnetostatic) interlayer coupling of 10 mT acting on Py and reproduce the low-frequency upturn in linewidth, seen to be present equally with and without MgO. \textit{Inset:} enhancements of nc-FMR linewidth over pc-FMR linewidth for the three samples.}
\label{fig3}
\end{figure}
\indent In order to determine whether the low-frequency broadening is related to spin pumping, we have also measured pc- and nc-FMR linewidths of Py(5 nm)/Cu(5 nm)/MgO(2 nm) and Py(5 nm)/Cu(3 nm)/MgO(2 nm)/Co(5 nm) structures, deposited with the same seed and capping layers. MgO interlayers are known to suppress spin pumping\cite{mosendzAPL2010}. Introducing MgO between Py and Co, we show in Fig. \ref{fig3} that the pc linewidths of Py in trilayer Py/Cu/Co (blue crosses) are restored to those of single-layer Py/Cu/MgO (overlapping green and red crosses), demonstrating suppression of spin pumping between Py and Co. However, we see a very similar upturn in low-frequency ($<10$ GHz) Py linewidth in nc-FMR (red circles), similar to that shown in Fig. \ref{fig2}(c). We attribute this low-frequency behavior to an interlayer coupling which cants the magnetization of Py a few degrees off the film normal when Co is not fully saturated along the film normal (i.e. $H_\text{B}<M_\text{eff}$). The solid curves in Fig. 3 assume a coupling field of 10 mT on Py, parallel to the local Co magnetization, which reproduce the linewidth broadening of nc-FMR. The peak-like features around 3 GHz show the maximal Gilbert damping enhancement when the Py magnetization is canted, as demonstrated in Fig. 5 \textit{inset}.\\
\indent Fig. 4 shows the central result of the paper. We compare the spin-pumping linewidth enhancements, $\mu_0(\Delta H^\text{tri}_{1/2}-\Delta H^\text{single}_{1/2})$, between pc- and nc-FMR (crosses and circles) in Fig. 4(a-d). Here $\Delta H^\text{single}_{1/2}$ and $\Delta H^\text{tri}_{1/2}$ are the linewidths of Py$_{1-x}$Cu$_x$ in Py$_{1-x}$Cu$_x$/Cu single layers and Py$_{1-x}$Cu$_x$/Cu/Co trilayers, respectively. The spin pumping linewidths are quite linear as a function of frequency for the pc-FMR data, as expected.  However, above 10 GHz (shaded regions), they are also quite linear in nc-FMR, which is not expected.  Collinear and noncollinear spin pumping linewidths agree closely.  This behavior is in contrast to the predicted behavior using $\lambda^L_\text{sr}=38$ nm for Co, measured by CPP-GMR\cite{pirauxEPJB1998}, and calculated in dashed curves according to the theory in the Supplemental Materials.  From the evident agreement between pc- and nc-linewidths above 10 GHz, for all Cu content $x$, we find no evidence for anisotropy in spin relaxation in our Co films.  Best fits to the data yield longitudinal spin relaxation lengths $\lambda^L_\text{sr}< 2$ nm in each of the four cases, approximately equal to the previously measured transverse length $\lambda_\text{sr}^T=2.4$ nm\cite{ghoshPRL2012}.\\
\begin{figure}[htb]
\centering
\includegraphics[width = 3.0 in] {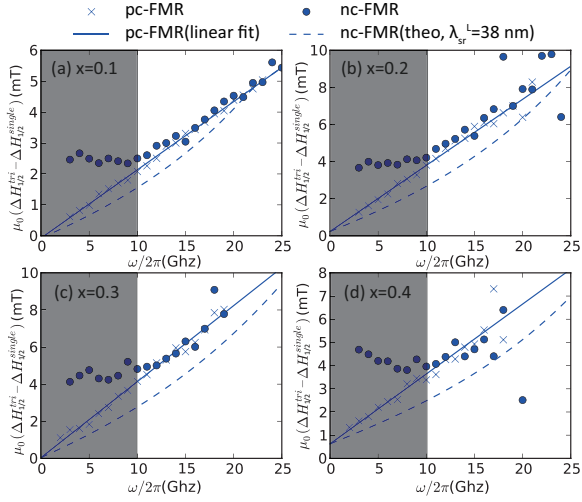}
\caption{Spin pumping contribution to linewidth in pc- and nc-FMR. (a-d) Linewidth enhancement of Py$_{1-x}$Cu$_x$ between single layers and trilayers in pc- and nc-FMR, $x=0.1$-0.4.
Solid lines are linear fits to the pc data (crosses); dashed curves are predicted from Eq. (1) using $\lambda^L_\text{sr}$=38 nm. The shadows at $\omega/2\pi\le 10$ GHz denote where the low-frequency linewidth broadenings are significant.}
\label{fig4}
\end{figure}
\indent Our model has assumed single-domain (macrospin) behavior in both Co and Py$_{1-x}$Cu$_x$ layers. For Py$_{1-x}$Cu$_x$ under field bias well in excess of $M_s$, the magnetization is well saturated, but for the Co layer, with higher $M_s$, nonuniform magnetization is possible. For greater control over the Co domain state, we have also carried out angle-dependent, fixed-frequency FMR measurements on Py$_{0.8}$Cu$_{0.2}$ and Py$_{0.8}$Cu$_{0.2}$/Cu/Co. Here the Co layer can be saturated more easily because the biasing field is canted away from the normal condition. The frequency is set to 10 GHz, where the low-frequency linewidth broadening of Py$_{0.8}$Cu$_{0.2}$ is insignificant (Figs. 3 and 4). As the field angle $\theta_\text{H}$ goes from 90$^\circ$ to 0$^\circ$ (pc to nc), the angle between the magnetizations of Py$_{0.8}$Cu$_{0.2}$ and Co changes from zero to maximum noncollinearity ($\sim50^\circ$) and $\Delta\alpha_\text{sp}$ would be expected to decrease significantly where the spin relaxation length in Co is markedly anisotropic.\\
\begin{figure}[htb]
\centering
\includegraphics[width = 3.0 in] {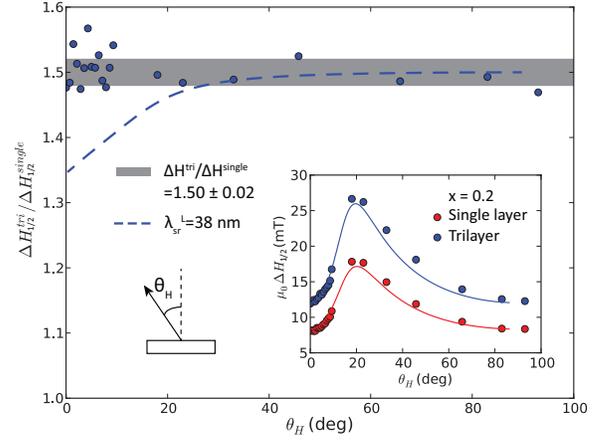}
\caption{Angle dependent linewidth ratio $\Delta H^\text{tri}_{1/2}/\Delta H^\text{single}_{1/2}$. The shadowed region shows the average with errorbar (1.50$\pm$0.02). \textit{Inset:} Angular dependence of $\mu_0 \Delta H_{1/2}$ for Py$_{0.8}$Cu$_{0.2}$ and Py$_{0.8}$Cu$_{0.2}$/Cu/Co at $\omega/2\pi=10$ GHz. Solid lines are macrospin calculations.}
\label{fig5}
\end{figure}
\indent Fig. \ref{fig5} \textit{Inset} shows the angular dependence of $\Delta H^\text{single}_{1/2}$ (red) and $\Delta H^\text{tri}_{1/2}$ (blue) for Py$_{0.8}$Cu$_{0.2}$. The data can be reproduced through a macrospin model\cite{platowPRB1998,mizukamiPRB2002} as shown in the solid curves, using similar magnetizations and {\it isotropic} dampings extracted from Fig. 2(a) and (c) ($\mu_0M_\text{eff}=0.53$ T, $\alpha_1=0.0114$ for the single layer, $\mu_0M_\text{eff}=0.55$ T, $\alpha_3=0.0168$ for the trilayer). The inhomogeneous broadenings are negligible, shown in Fig. 2(c). For small enough $\theta_\text{H}$, the resonance field of the Co starts to fall below the expected macrospin value, as shown in the Supplemental Materials, Section C\cite{supplemental}. We take the angle at which this behavior appears (at $\theta_\text{H} \sim 18^\circ$) to be the limit above which we have the greatest confidence in single-domain ordering of Co.\\
\indent In the main panel of Fig. \ref{fig5} we replot the trilayer and single-layer linewidths for Py$_{0.8}$Cu$_{0.2}$, shown in the inset, as the ratio $\Delta H^\text{tri}_{1/2}/\Delta H^\text{single}_{1/2}$. Because the inhomogeneous linewidths are negligible for the structures ($< 0.5$ mT), the linewidth ratio for {\it isotropic} spin pumping would be approximated well through the ratio of the Gilbert damping for the two configurations, $\Delta H^\text{tri}_{1/2}/\Delta H^\text{single}_{1/2} =  1+\Delta\alpha_\text{sp}/\alpha_1$.  We find that the linewidth ratio is in fact constant within experimental error, shown by the shaded region in Fig. \ref{fig5}. The blue dashed curve shows the expected behavior for anisotropic spin relaxation, assuming $\lambda_\text{sr}^L= 38$ nm, with a marked decrease in the linewidth ratio for low angles $\theta_\text{H}$.  A best fit to these data returns $\lambda_\text{sr}^L< 1.1$ nm.  If we restrict our attention to field angles $\theta_\text{H} \geq \textrm{18}^\circ$, above which we have confidence in macrospin behavior of the Co layer, the best fit is not changed greatly, with $\lambda_\text{sr}^L\leq 4$ nm, within experimental error of the transverse length $\lambda_\text{sr}^T$.\\
\indent Extrinsic effects, i.e. issues of sample quality, may play some role in the results. First, longitudinal spin relaxation lengths $\lambda_\text{sr}^L$, if equated with the spin diffusion length $\lambda_\text{sd}$, are inversely proportional to (defect-related) resistivity\cite{jbassJMMM1999}. However, four-point probe measurements of the resistivity of our Co (5 nm) films show 25 $\mu\Omega\cdot$cm, comparable with the 18 $\mu\Omega\cdot$cm reported in the room-temperature CPP-GMR experiment\cite{pirauxEPJB1998}, and therefore comparably long spin diffusion lengths should be expected. Second, we see that the spin mixing conductances $g^{\uparrow\downarrow}_\text{eff}$ of Py$_{1-x}$Cu$_x$/Cu/Co measured here are lower than those measured in Ref. \cite{ghoshPRL2012}, on structures deposited elsewhere.  The most plausible source of the reduction is a more resistive Cu layer, which adds an additional resistive term\cite{zwierzyckiPRB2005,ghoshAPL2011} $(2e^2/h)t_\text{Cu}\rho_\text{Cu}$ to $g^{\uparrow\downarrow}_\text{eff}$. Here however the bulk Cu properties should have little influence over either spin relaxation length and should not affect the anisotropy of spin relaxation strongly.\\
\indent Our estimate of $\lambda^L_\text{sr}$ in Co is consistent with a general observation that spin relaxation as measured in spin pumping/FMR is shorter-ranged than it is as measured in magnetotransport. In Pd and Pt, the characteristic relaxation lengths for dynamically pumped spin current are measured as 1-5 nm\cite{ghoshPRL2012, zhangAPL2013, vlaminckPRB2013,CaminalePRB2016}, whereas in GMR they are closer to 10-20 nm\cite{kurtAPL2002, morotaPRB2011}. We suggest therefore that the quantities revealed by the two types of measurements may differ in some respect. For example, robust spin-pumping effects have been found in ferrimagnetic insulators such as yttrium iron garnet (YIG). These effects clearly have little to do with electronic transport in YIG, and their characteristic lengths would refer to scattering mechanisms distinct from those involved in CPP-GMR. A second possibility, alluded to in the review in Ref. \cite{jbassJPCM2007}, is that the room-temperature spin diffusion length of 38 nm in \cite{pirauxEPJB1998} is an overestimate due to technical issues of the CPP-GMR measurement in Co multilayers; the majority of such measurements in various ferromagnets show $<10$ nm\cite{jbassJPCM2007}. Our results, in this scenario, may alternately imply that the short spin diffusion length observed in Py is not far away from that of Co.\\
\indent In summary, we have experimentally demonstrated that the spin relaxation in Co, as measured by noncollinear spin pumping, is largely isotropic. The estimated longitudinal spin relaxation length, $<$ 2 nm, is an order of magnitude smaller than measured by magnetotransport but comparable to the transverse spin relaxation length. We acknowledge NSF-DMR-1411160 for support.\\
\bibliographystyle{ieeetr}

\begin{thebibliography}{5}
\bibitem{anglealignment}
In nc-FMR, the sample normal is aligned carefully (about two axes, with $<$0.2$^\circ$ precision) to maximize $\mu_0H_\text{res}$ of the Py$_{1-x}$Cu$_x$ layers at 3 GHz. This step is critical to reduce the inhomogeneous broadening.
\end{thebibliography}

\end{document}




\title{Supplemental Material to "Characterization of spin relaxation anisotropy in Co using spin pumping"}

\author{Y. Li, W. Cao and W. E. Bailey}

\affiliation{Dept. of Applied Physics \& Applied Mathematics, Columbia University, New York NY 10027, USA}

\date{\today}

\maketitle
\subsection{Calculation of spin-pumping damping for noncollinearly magnetized, asymmetric trilayers}
\begin{figure}[htb]
\centering
\includegraphics[width = 5.0 in] {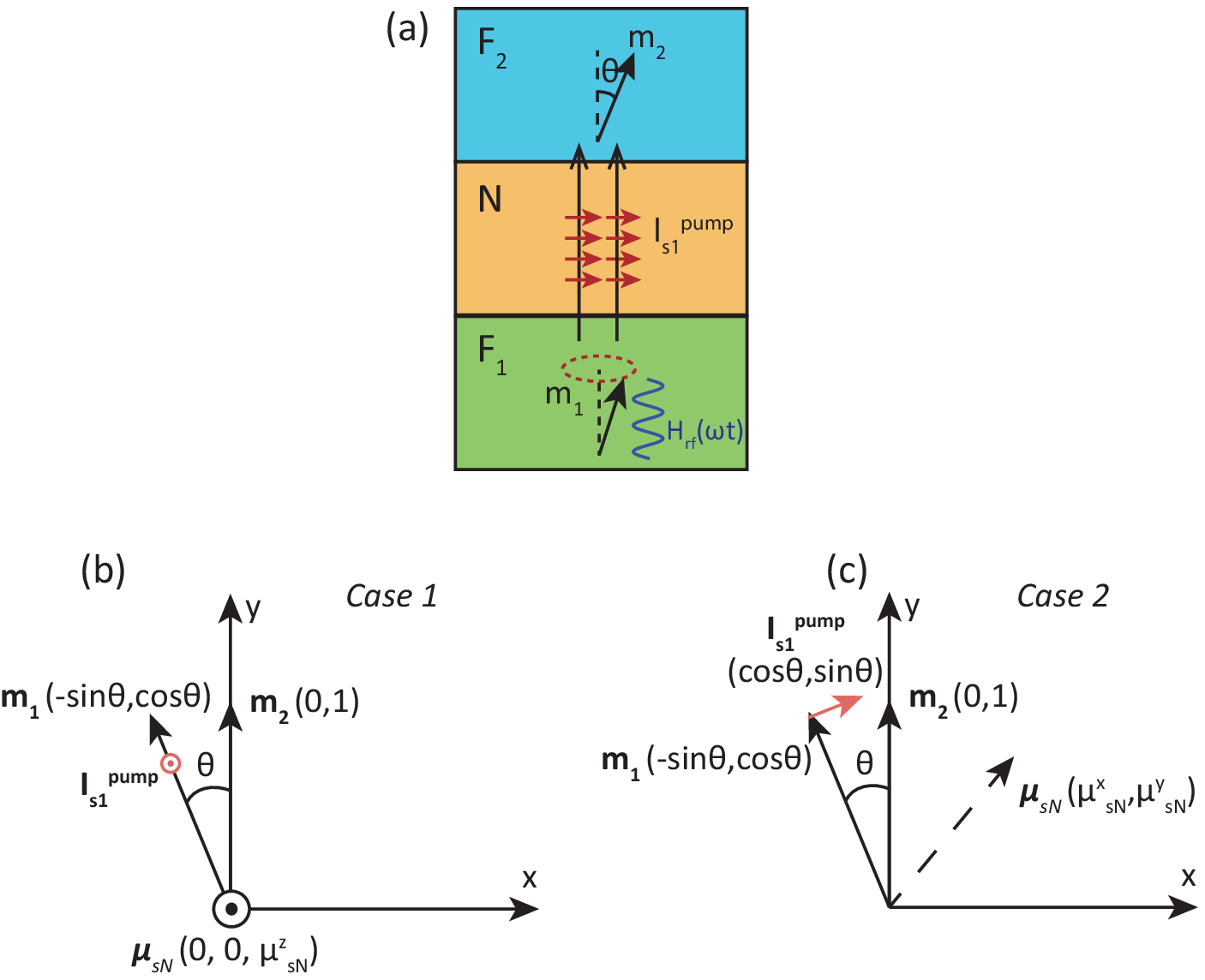}
\caption{(a) Magnetization configuration of the asymmetric F$_1$/N/F$_2$ trilayer. (b) An instant in which the spin polarization of I$_\text{s1}^\text{pump}$ is orthogonal to both \textbf{m}$_1$ and \textbf{m}$_2$. $\boldsymbol{\mu}_{sN}$ is also orthogonal to \textbf{m}$_1$ and \textbf{m}$_2$. (c) An instant in which the spin polarization of $\boldsymbol{I}_\text{s1}^\text{pump}$ is in the same plane of \textbf{m}$_1$ and \textbf{m}$_2$. $\boldsymbol{\mu}_{sN}$ is also in the same plane of \textbf{m}$_1$ and \textbf{m}$_2$}
\end{figure}

\indent Consider an asymmetric ferromagnet / noble metal / ferromagnet (F$_1$/N/F$_2$) spin-valve trilayer structure, shown in Fig. 1(a).  The time-averaged magnetization of F$_1$ is pictured along the film-normal, although it can take any angle with respect to the film-normal. We assume that F$_1$ undergoes small-angle precession. F$_2$ is noncollinearly magnetized with respect to F$_1$, where $\theta$ is the  angle of noncollinearity or misalignment; $\theta=0$ for parallel magnetizations $\mathbf{m}_1=\mathbf{m}_2$, where $\mathbf{m}_i$ , $i=1$ or 2, is the unit vector of the magnetization $\mathbf{M}_i$ of F$_i$. The magnetizaton of F$_2$ is taken to be stationary. The spin current flows from the N spacer to each of the F layers F$_1$, F$_2$ are\cite{tserkovnyakPRB2003,tserkovnyakRMP2005,taniguchiPRB2007,taniguchiAPEX2008}:
\begin{equation}
\mathbf{I}_{s}^{\text{N}\rightarrow\text{F}_1} = {\
g^*_1 \over 4\pi} \mathbf{m}_1(\boldsymbol{\mu}_{sN}\cdot\mathbf{m}_1) \
+ {\tilde{g}_1^{\uparrow\downarrow}\over 4\pi} \mathbf{m}_1\times \boldsymbol{\mu}_{sN} \times\mathbf{m}_1
\label{eq01}
\end{equation}
\begin{equation}
\mathbf{I}_{s}^{\text{N}\rightarrow\text{F}_2} = {\
g^*_2 \over 4\pi} \mathbf{m}_2(\boldsymbol{\mu}_{sN}\cdot\mathbf{m}_2) \
+ {\tilde{g}_2^{\uparrow\downarrow}\over 4\pi} \mathbf{m}_2\times \boldsymbol{\mu}_{sN} \times\mathbf{m}_2
\label{eq02}
\end{equation}
where $\boldsymbol{\mu}_{sN}$ is the spin accumulation vector in the N layer, $g_i^*$ and $\tilde{g}_i^{\uparrow\downarrow}$ are the effective longitudinal spin conductance and transverse spin mixing conductance for F$_i$/N interface, respectively. Here the spin current vector denotes the direction of spin polarization, the direction of current flow always being normal to interfaces.
The conservation of spin angular momentum, assuming spin-current conservation (negligible dissipation) N, gives:
\begin{equation}
\mathbf{I}_{s}^{\text{N}\rightarrow\text{F}_1} + \mathbf{I}_{s}^{\text{N}\rightarrow\text{F}_2} = \mathbf{I}_{s1}^\text{pump}
\label{eq03}
\end{equation}
where $\mathbf{I}_{s1}^\text{pump}$ is the pumped spin current from F$_1$ into N\cite{tserkovnyakPRL2002,tserkovnyakRMP2005}:
\begin{equation}
\mathbf{I}_{s1}^\text{pump} = {\hbar \over 4\pi} \tilde{g}_1^{\uparrow\downarrow} \mathbf{m}_1\times \mathbf{\dot{m}_1}
\label{eq04}
\end{equation}
Substituting Eq. (\ref{eq01}), (\ref{eq02}) and (\ref{eq04}) into the continuity expression (\ref{eq03}), we obtain a vector equation in terms of the vector spin accumulation $\boldsymbol{\mu}_{sN}$. To calculate the spin pumping damping enhancement, we seek solutions for $\boldsymbol{\mu}_{sN}$ in order to find the spin current flow into $\mathbf{m}_2$, which is absorbed by $\mathbf{m}_2$.\\
\indent The vector $\mathbf{I}_{s1}^\text{pump}$, proportional to $\mathbf{m}_1\times \mathbf{\dot{m}_1}$, rotates in the plane with normal given by $\mathbf{m}_1$. Assuming a finite misalignment angle $\theta$ between $\mathbf{m}_1$ and $\mathbf{m}_2$, $\mathbf{I}_{s1}^\text{pump}$ will oscillate between fully orthogonal to $\mathbf{m}_2$ (Fig. 1b) and canted away from orthogonality by $\theta$ (Fig. 1c). We consider these two extreme cases during the precession of $\mathbf{m}_1$. In case 1 (Fig. 1b), $\mathbf{I}_{s1}^\text{pump}$ is perpendicular to both $\mathbf{m}_1$ and $\mathbf{m}_2$. In case 2 (Fig. 1c), $\mathbf{I}_{s1}^\text{pump}$ is in the same plane as $\mathbf{m}_1$ and $\mathbf{m}_2$ and has the largest longitudinal component along $\mathbf{m}_2$.\\
\indent In \textit{case 1}, $\mathbf{I}_{s}^{\text{N}\rightarrow\text{F}_1}$, $\mathbf{I}_{s}^{\text{N}\rightarrow\text{F}_2}$ and $\boldsymbol{\mu}_{sN}$ are all parallel to $\mathbf{I}_{s1}^\text{pump} $. In Eqs. (\ref{eq01}) and (\ref{eq02}) the first terms become zero and only the second terms remain. The solution of Eqs. (\ref{eq01})-(\ref{eq03}) has a scalar form along the direction $\hat{z}$ orthogonal to both F$_1$ and F$_2$:
\begin{equation}
I_{s,z}^{\text{N}\rightarrow\text{F}_1\{\text{F}_2\}} =\left({\tilde{g}_{1\{2\}}^{\uparrow\downarrow} \over \tilde{g}_1^{\uparrow\downarrow} + \tilde{g}_2^{\uparrow\downarrow}}\right) I_{s1}^\text{pump},\,\,\,\, {\mu}_{sN}^z = \left({4\pi \over \tilde{g}_1^{\uparrow\downarrow} + \tilde{g}_2^{\uparrow\downarrow}}\right) I_{s1}^\text{pump}
\label{eq05}
\end{equation}
It has been shown previously that\cite{tserkovnyakPRL2002,tserkovnyakRMP2005} the dissipation of spin angular momentum due to a transverse spin mixing conductance $g^{\uparrow\downarrow}$ leads to an additional Gilbert damping term $\Delta\alpha = {\gamma \hbar g^{\uparrow\downarrow} / 4\pi M_{s}d}$. With only $\mathbf{I}_{s}^{\text{N}\rightarrow\text{F}_2}$ dissipated, the spin-pumping damping enhancement can be expressed as:
\begin{equation}
\Delta \alpha_{sp} = I_{s2}^\text{back} \cdot {\Delta\alpha_0 \over I_{s1}^\text{pump}} = \Delta\alpha_0 \cdot {\tilde{g}_2^{\uparrow\downarrow} \over \tilde{g}_1^{\uparrow\downarrow} + \tilde{g}_2^{\uparrow\downarrow}}
\label{eq06}
\end{equation}
with $\Delta\alpha_0 = {\gamma \hbar \tilde{g}_1^{\uparrow\downarrow} / 4\pi M_{s1}d}$. Eq. (\ref{eq06}) is identical to the collinear spin pumping case with an effective spin mixing conductance $(\tilde{g}_{eff}^{\uparrow\downarrow})^{-1} = (\tilde{g}_1^{\uparrow\downarrow})^{-1} + (\tilde{g}_2^{\uparrow\downarrow})^{-1}$.\\
\indent In \textit{case 2}, $\boldsymbol{\mu}_{sN}$ has only a component coplanar with $\mathbf{m}_1$ and $\mathbf{m}_2$ ($\mu_{sN}^x$ and $\mu_{sN}^y$). In Eq. (\ref{eq01}) and (\ref{eq02}) both terms need to be considered. The $\hat{x}$ and $\hat{y}$ components of Eq. (\ref{eq03}) can be written as:
\begin{align}
4\pi I_{s1}^\text{pump}\binom{\cos\theta}{\sin\theta} = &g_1^*(-\mu_{sN}^x\sin\theta+\mu_{sN}^y\cos\theta)\binom{-\sin\theta}{\cos\theta} + \tilde{g}_1^{\uparrow\downarrow}(\mu_
{sN}^x\cos\theta+\mu_{sN}^y\sin\theta)\binom{\cos\theta}{\sin\theta} \nonumber \\
+ &g_2^*\mu_{sN}^y \binom{0}{1} + \tilde{g}_2^{\uparrow\downarrow}\mu_{sN}^x \binom{1}{0}
\label{eq07}
\end{align}
The solution of Eq. (\ref{eq07}) can be expressed as:
\begin{subequations}
 \begin{eqnarray}
  \mu_{sN}^x = {4\pi I_{s1}^\text{pump}(C \cos\theta-B \sin\theta) \over AC-B^2}\\
  \mu_{sN}^y = {4\pi I_{s1}^\text{pump}(A \sin\theta-B \cos\theta) \over AC-B^2}
 \end{eqnarray}
 \label{eq08}
\end{subequations}
where
\begin{subequations}
 \begin{eqnarray}
  A(\theta) = &g_1^*\sin^2\theta+\tilde{g}_1^{\uparrow\downarrow}\cos^2\theta + \tilde{g}_2^{\uparrow\downarrow}\\
  B(\theta) = &(\tilde{g}_1^{\uparrow\downarrow} - g_1^*)\sin\theta\cos\theta\\
  C(\theta) = &g_1^*\cos^2\theta+\tilde{g}_1^{\uparrow\downarrow}\sin^2\theta + g_2^*
 \end{eqnarray}
 \label{eq09}
\end{subequations}
The spin torque is equal to the conponent of $\mathbf{I}_{s}^{\text{N}\rightarrow\text{F}_2}$ transverse to $\mathbf{m}_1$, or the component which is parallel to $\mathbf{I}_{s1}^\text{pump}$. Thus the spin-pumping damping enhancement can be written in terms of the defined misalignment-dependent quantities $A(\theta)$, $B(\theta)$, $C(\theta)$ as:
\begin{align}
\Delta \alpha_{sp}(\theta) =& {\mathbf{I}_{s}^{\text{N}\rightarrow\text{F}_2}\cdot \mathbf{I}_{s1}^\text{pump} \over I_{s1}^\text{pump}} \cdot {\Delta\alpha_0 \over I_{s1}^\text{pump}} \nonumber \\
=& \Delta\alpha_0 \cdot {g_2^*(A\sin^2\theta-B\sin\theta\cos\theta)+\tilde{g}_2^{\uparrow\downarrow}(C\cos^2\theta-B\sin\theta\cos\theta) \over AC-B^2}
\label{eq10}
\end{align}
It is easy to verify that at $\theta=0^\circ$ Eq. (\ref{eq10}) recovers Eq. (\ref{eq06}), same as the collinear spin pumping.\\
\indent Having treated the two special spin current orientations, we need to take the average of all the orientation possibilities. We refer to the calculation by Taniguchi, et al.\cite{taniguchiPRB2007}, that in a symmetric spin valve (F$_1$=F$_2$) the small-precession limit of averaged spin-pumping damping enhancement is equal to the arithmetic mean of damping enhancement with out-of-plane $\mathbf{I}_{s1}^\text{pump}$ (case 1) and in-plane $\mathbf{I}_{s1}^\text{pump}$ (case 2). The Eq. 13 in Ref. \cite{taniguchiPRB2007} can be simplified, at small precession angle, as:
\begin{equation}
\Delta\alpha_{sp} = \Delta\alpha_0 \
\left[ 1-{(\nu/2) \sin^2\theta \over {1-\nu^2\cos^2\theta} } \right]
\label{eq11}
\end{equation}
which is the average of $\Delta\alpha_0$ and $\Delta\alpha_0[1-\nu\sin^2\theta/(1-\nu^2\cos^2\theta)]$ (Eq. 5 in Ref. \cite{taniguchiPRB2007}). We apply it to the asymmetric spin valve condition: all the theoretical curves in the main text are calculated from the mean of Eq. (\ref{eq06}) and Eq. (\ref{eq10}). \\
\indent The theoretical curves in Fig. 4 and 5 of the main text are calculated using the routine, assuming $\lambda^L_{sr}=38$ nm for Co. The new estimation of $\lambda^L_{sr}$ ($<2$ nm) in the manuscript takes the best value that fits the damping calculation to the experimental data.\\

\subsection{Values of $g^*$ and $\tilde{g}^{\uparrow\downarrow}$}
\indent In this section we calculate the value of the two effective spin conductances. The transverse spin mixing conductance $\tilde{g}^{\uparrow\downarrow}$ (Sharvin correction includedtserkovnyakRMP2005) of each interface can be calculated from the effective spin mixing conductance of Py$_{1-x}$Cu$_x$/Cu/Co structures and the comparison measurements of Py$_{1-x}$Cu$_x$/Cu/Pt and Co/Cu/Pt (Table \ref{table1}). For Py$_{1-x}$Cu$_x$/Cu/Co, the total spin mixing conductance can be expressed as:
\begin{equation}
{1 \over g^{\uparrow\downarrow}_{\text{Py}_{1-x}\text{Cu}_x/\text{Cu}/\text{Co}}} = {1 \over \tilde{g}^{\uparrow\downarrow}_{\text{Py}_{1-x}\text{Cu}_x/\text{Cu}}}+{1 \over \tilde{g}^{\uparrow\downarrow}_{\text{Co}/\text{Cu}}}
\label{eq12}
\end{equation}
For F/Cu/Pt (F=Py$_{1-x}$Cu$_x$ or Co), the effective spin mixing conductance can be formulated as:
\begin{equation}
{1 \over g^{\uparrow\downarrow}_{\text{F}/\text{Cu}/\text{Pt}}} = {1 \over \tilde{g}^{\uparrow\downarrow}_{\text{F}/\text{Cu}}}+{1 \over \tilde{g}^{\uparrow\downarrow}_{\text{Cu}/\text{Pt}}}
\label{eq13}
\end{equation}
In the experiment the thicknesses of Pt are kept the same and we can treat $\tilde{g}^{\uparrow\downarrow}_{\text{Cu}/\text{Pt}}$ as a constant. Solving Eq. (\ref{eq12}) and (\ref{eq13}) we obtain:
\begin{subequations}
 \begin{eqnarray}
  {1 \over \tilde{g}^{\uparrow\downarrow}_{\text{Py}_{1-x}\text{Cu}_x/\text{Cu}}} = {1 \over g^{\uparrow\downarrow}_{\text{Py}_{1-x}\text{Cu}_x/\text{Cu}/\text{Co}}} + {1 \over g^{\uparrow\downarrow}_{\text{Py}_{1-x}\text{Cu}_x/\text{Cu}/\text{Pt}}} - {1 \over g^{\uparrow\downarrow}_{\text{Co}/\text{Cu}/\text{Pt}}}\\
  {1 \over \tilde{g}^{\uparrow\downarrow}_{\text{Co}/\text{Cu}}} = {1 \over g^{\uparrow\downarrow}_{\text{Py}_{1-x}\text{Cu}_x/\text{Cu}/\text{Co}}} - {1 \over g^{\uparrow\downarrow}_{\text{Py}_{1-x}\text{Cu}_x/\text{Cu}/\text{Pt}}} + {1 \over g^{\uparrow\downarrow}_{\text{Co}/\text{Cu}/\text{Pt}}}
 \end{eqnarray}
 \label{eq14}
\end{subequations}

In Table \ref{table2} we list the calculated values of $\tilde{g}^{\uparrow\downarrow}_{\text{Py}_{1-x}\text{Cu}_x/\text{Cu}}$ and $\tilde{g}^{\uparrow\downarrow}_{\text{Co}/\text{Cu}}$. For $x=0.1$ and 0.3 we take the linear interpolated values to evaluate $g^{\uparrow\downarrow}_{\text{Py}_{1-x}\text{Cu}_x/\text{Cu}/\text{Pt}}$. In addition, we also show the values compensating the Sharvin correction, with $1/g^{\uparrow\downarrow}_i = 1/\tilde{g}^{\uparrow\downarrow}_i+1/2g^\text{Sh}_\text{Cu}$, $g^\text{Sh}_\text{Cu}=15$ nm$^{-2}$.

Compared with previous measurements\cite{ghoshPRL2012,ghoshAPL2011}, we find smaller values of $g^{\uparrow\downarrow}_{\text{Py}_{1-x}\text{Cu}_x/\text{Cu}/\text{Co}}$ and $g^{\uparrow\downarrow}_{\text{Py}_{1-x}\text{Cu}_x/\text{Cu}}$ for $x=0$. However we argue that the spin mixing conductances of Co/Cu/Pt in Table \ref{table1} and Co/Cu interfaces in Table \ref{table2} are reasonable, which ensures a good Co/Cu interface crucial for the study of spin relaxation anisotropy. It is also possible that a resistive Cu spacer contributes an additional resistance, $(2e^2/h)t_\text{Cu}\rho_\text{Cu}$\cite{tserkovnyakRMP2005}, to the right side of Eq. (\ref{eq13}). To reduce the spin mixing conductance of Py/Cu/Co from 15.0 nm$^{-2}$ in Ref. \cite{ghoshPRL2012} to 7.6 nm$^{-2}$ in Table \ref{table1}, one needs to take $\rho_\text{Cu}=16.8$ $\mu\Omega\cdot$cm. However we point out that this resistive scattering will contribute to both transverse and longitudinal spin conductance by the same amount, and the anisotropy of spin relaxation should not be affected. In practice, we take the effective interfacial spin mixing conductance into the model for the estimation of $\lambda_{sr}^L$ and and use the values of $\lambda_{sr}^T$ from Ref. \cite{ghoshPRL2012}.

\begin{center}
\begin{table*}[htb]
\centering
\begin{tabular}{>{\centering\arraybackslash}m{1.2in} | >{\centering\arraybackslash}m{0.6in} >{\centering\arraybackslash}m{0.6in} >{\centering\arraybackslash}m{0.6in} >{\centering\arraybackslash}m{0.6in} >{\centering\arraybackslash}m{0.6in} }
\hline
(Unit: nm$^{-2}$) & $x=0$ & $x=0.1$ & $x=0.2$ & $x=0.3$ & $x=0.4$ \\
\hline
\hline
$g^{\uparrow\downarrow}_{\text{Py}_{1-x}\text{Cu}_x/\text{Cu}/\text{Co}}$ & 7.6 & 5.6 & 7.3 & 6.8 & 6.8\\
$g^{\uparrow\downarrow}_{\text{Py}_{1-x}\text{Cu}_x/\text{Cu}/\text{Pt}}$ & 6.0 & - & 5.0 & - & 5.0\\
$g^{\uparrow\downarrow}_{\text{Co}/\text{Cu}/\text{Pt}}$ & 7.9 &  & &  & \\
\hline
\end{tabular}
\caption{Experimental values of (effective) spin mixing conductance of $\text{Py}_{1-x}\text{Cu}_x/\text{Cu}/\text{Co}$, $\text{Py}_{1-x}\text{Cu}_x/\text{Cu}/\text{Pt}$ and $\text{Co}/\text{Cu}/\text{Pt}$ samples, extracted from spin-pumping linewidth enhancements. }
\label{table1}
\end{table*}
\end{center}

\begin{center}
\begin{table*}[ht]
\centering
\begin{tabular}{>{\centering\arraybackslash}m{1.2in} | >{\centering\arraybackslash}m{0.6in} >{\centering\arraybackslash}m{0.6in} >{\centering\arraybackslash}m{0.6in} >{\centering\arraybackslash}m{0.6in} >{\centering\arraybackslash}m{0.6in} }
\hline
(Unit: nm$^{-2}$) & $x=0$ & $x=0.1$ & $x=0.2$ & $x=0.3$ & $x=0.4$ \\
\hline
\hline
$\tilde{g}^{\uparrow\downarrow}_{\text{Py}_{1-x}\text{Cu}_x/\text{Cu}}$ & 11.7 & 8.6 & 9.5 & 9.1 & 9.1\\
$\tilde{g}^{\uparrow\downarrow}_{\text{Co}/\text{Cu}}$ & 21.9 & 16.2 & 31.5 & 27.2 & 27.2\\
\hline
$g^{\uparrow\downarrow}_{\text{Py}_{1-x}\text{Cu}_x/\text{Cu}}$ & 8.4 & 6.7 & 7.2 & 7.0 & 7.0\\
$g^{\uparrow\downarrow}_{\text{Co}/\text{Cu}}$ & 12.7 & 10.5 & 15.4 & 14.3 & 14.3\\
\hline
\end{tabular}
\caption{``$\tilde{g}^{\uparrow\downarrow}_i$'': Sharvin-corrected spin mixing conductance of $\text{Py}_{1-x}\text{Cu}_x/\text{Cu}$ and $\text{Co}/\text{Cu}$ interfaces, calculated from Eq. (\ref{eq14}). ``$g^{\uparrow\downarrow}_i$'': interfacial spin mixing conductance compensating the Sharvin conductance of Cu layer. $i=$ Py$_{1-x}$Cu$_x$/Cu or Co/Cu.}
\label{table2}
\end{table*}
\end{center}

\indent The effective longitudinal spin conductance $g^*$ can be expressed as\cite{tserkovnyakPRB2003}:
\begin{equation}
{1\over g^*} = {{g^{\uparrow\uparrow}+g^{\downarrow\downarrow}}\over 2g^{\uparrow\uparrow}g^{\downarrow\downarrow}} \
+ {1\over {g_{sd}\tanh(t_{F}/\lambda^L_{sr})}}
\label{eq15}
\end{equation}
In the first term, $g^{\uparrow\uparrow\{\downarrow\downarrow\}}_i$ is the interfacial spin-up \{spin-down\} conductance. $g^{\uparrow\uparrow\{\downarrow\downarrow\}}$ can be calculated by $1/g^{\uparrow\uparrow\{\downarrow\downarrow\}} = (e^2/h) AR^{\uparrow\{\downarrow\}}_{\text{F/N}}$ where $AR^{\uparrow\{\downarrow\}}_{\text{F/N}}$ is the electron interface resistance. We use the experimental value from GMR measurements: $2AR^* = (AR^{\uparrow}+AR^{\downarrow})/2=1.04$ f$\Omega\cdot$m$^2$ for Co/Cu\cite{jbassIEEE1998} and 1.0 f$\Omega\cdot$m$^2$ for Py/Cu\cite{steenwykJMMM1997}. We can calculate that $2g^{\uparrow\uparrow}g^{\downarrow\downarrow}/(g^{\uparrow\uparrow}+g^{\downarrow\downarrow})=26$ nm$^{-2}$ for both interfaces. \\
\indent In the second term, $g_{sd}$ has been expressed in Ref. \cite{tserkovnyakPRB2003} as:
\begin{equation}
g_{sd} = {h \over e^2\lambda^L_{sr}} {2\sigma^\uparrow\sigma^\downarrow \over \sigma^\uparrow+\sigma^\downarrow}
\label{eq16}
\end{equation}
where $\sigma^{\uparrow, \downarrow}$ are the spin-up/down electron conductivity in F, $h$ is the Planck constant and $e$ is the electronic charge. Here we simply take $\sigma^\uparrow=\sigma^\downarrow=\sigma/2$ ($\sigma$ is the total electrical conductivity), which has also been done in Eq. (74) of Ref. \cite{tserkovnyakRMP2005}. Following this treatment, the term $2\sigma^\uparrow\sigma^\downarrow / (\sigma^\uparrow+\sigma^\downarrow)$ is replaced by $\sigma/2$. Taking $\rho_{Co}=25$ $\mu\Omega\cdot$cm and $\rho_{Py}=30$ $\mu\Omega\cdot$cm from our four-point probe measurements and $\lambda^L_{sr}=38$ nm for Co\cite{pirauxEPJB1998} and 4.3 nm for Py\cite{duboisPRB1999} from the literatures, we calculate $g_{sd}\tanh(t_{F}/\lambda^L_{sr})$ to be 0.18 nm$^{-2}$ for Co and 8.3 nm$^{-2}$ for Py when the F thickness is 5 nm; the large disagreement comes from the expected difference in $\lambda^L_{sr}$. As a result, $g^*=0.18$ nm$^{-2}$ for Co and 6.2 nm$^{-2}$ for Py are obtained from Eq. (\ref{eq15}) and used to produce the theoretical curves in the manuscript.\\
\indent In the experiment, we do not find the anisotropic response of spin pumping predicted above. According to our model, the lack of anisotropic response can be explained best through a difference in the longitudinal spin conductance $g^*$ for Co/Cu, as this is the most dominant term in Eq. (\ref{eq10}) and sensitive to $\lambda_{sr}^L$. This is because in the experiments we choose the thickness of Co to be much less than 38 nm in order to examine the spin relaxation anisotropy. \\
\indent From Py to Py$_{1-x}$Cu$_x$, we should expect that both $g^{\uparrow\uparrow\{\downarrow\downarrow\}}$ and $\sigma^{\uparrow\{\downarrow\}}$ will increase due to a better conducting ability of Cu than Py. $\lambda^L_{sr}$ may also vary. However we emphasize that in Eq. (\ref{eq10}), the anisotropy is dominated by $g^*_2$ and $\tilde{g}^{\uparrow\downarrow}_2$ and not sensitive to $g^*_1$. For example in the angular dependence of linewidth ration for $x=0.2$ (Fig. 5 of the main text), increasing $g^*_{\text{Py}_{0.8}\text{Cu}_{0.2}/\text{Cu}}$ by a factor of two will change the single-domain estimation of $\lambda_{sr}^L$ from 1.8$\pm$2.7 to 2.1$\pm$2.8, still much smaller than the GMR measurements. Thus for simplicity we keep using the value of $g^*$ of Py for Py$_{1-x}$Cu$_x$ layers.\\

\subsection{Single-domain limit determination}
\begin{figure}[htb]
\centering
\includegraphics[width = 5.0 in] {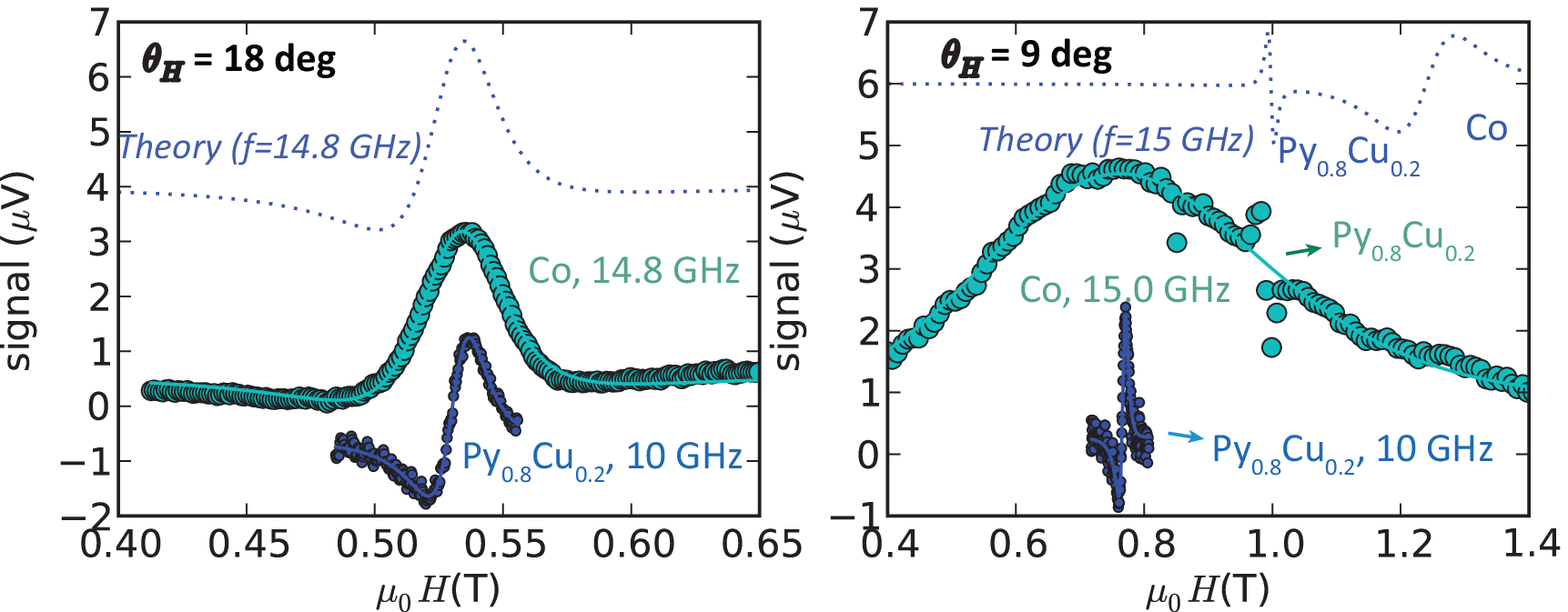}
\caption{Resonance peak of Co and Py$_{0.8}$Cu$_{0.2}$ independently measured in Py$_{0.8}$Cu$_{0.2}$/Cu/Co with  $\theta_H=18^\circ$ (a) and $9^\circ$ (b). The resonance frequency of Py$_{0.8}$Cu$_{0.2}$ are both 10 GHz. The resonance frequency of Co is adjusted so that the $\mu_0H_\text{res}$ of Co is equal to Py$_{0.8}$Cu$_{0.2}$. Dashed curves show the theoretical prediction of Co resonance signals.}
\end{figure}
\indent To determine whether the Co layer is in a single-domain state in Py$_{0.8}$Cu$_{0.2}$/Cu/Co when the Py$_{0.8}$Cu$_{0.2}$ layer is at resonance, we have measured the FMR signal of Co at different $\theta_H$. First we measure the FMR signal of Py$_{0.8}$Cu$_{0.2}$ at one angle and determine the resonance field $\mu_0 H_\text{res}$. Next we adjust the frequency so that the Co FMR signal can be measured at the same field. Then we compare the lineshape with the macrospin model prediction\cite{mizukamiJJAP2001}. In Fig. 2(a), when $\theta_H=18^\circ$ the resonance field $\mu_0 H_\text{res}$ for Py$_{0.8}$Cu$_{0.2}$ is 0.53 T at 10 GHz. For Co, the resonance field is located at 0.53 T for 14.8 GHz. The macrospin model for angle-dependent FMR shows $\omega/2\pi=14.8$ GHz, identical to the experiment, showing the Co can indeed be treated as a macrospin. However for $\theta_H=9^\circ$ (Fig. 2b), we find that the Co resonance is located at 15.0 GHz, quite different from the macrospin prediction of 12.3 GHz. To see the difference more clearly, we plot (dashed lines) the macrospin prediction for both Py$_{0.8}$Cu$_{0.2}$ and Co resonances at 15.0 GHz, based on the magnetizations and linewidths measured from perpendicular FMR. The Py$_{0.8}$Cu$_{0.2}$ peak matches with experiment. The calculated Co peak deviates from experiment in both resonance field and linewidth. Thus we determine the single-domain limit of $\theta_H$ to be somewhere between $9^\circ$ and $18^\circ$ in the sample. The upper bound $18^\circ$ is used in the manuscript for the single-domain limit.\\
\bibliographystyle{ieeetr}

